# Extraordinary Transmission in the UV Range from Sub-wavelength Slits on Semiconductors

M. A. Vincenti[1], D. de Ceglia[1], N. Akozbek[2], M. Buncick[1], M. J. Bloemer[2], M. Scalora[2]*

*1 AEgis Technologies Group, 631 Discovery Dr., Huntsville, AL 35806*
*2 Charles M. Bowden Research Center, AMSRD-AMR-WS-ST, RDECOM, Redstone Arsenal, Alabama 35898-5000,USA*

*\*Corresponding author: michael.scalora@us.army.mil*

**Abstract**: In this paper we describe a way to achieve the extraordinary transmission regime from sub-wavelength slits carved on semiconductor substrates. Unlike metals, the dielectric permittivity of typical semiconductors like GaAs or GaP is negative beginning in the extreme UV range ($\lambda \leq 270$nm). We show that the metal-like response of bulk semiconductors exhibits surface plasmon waves that lead to extraordinary transmission in the UV and soft X-ray ranges. The importance of realistic material response versus perfect conductors is also discussed. These findings may be important in high resolution photo-lithography, near field optical devices and ultra high density optical storage.

**References and links**

1. H.A. Bethe, "Theory of diffraction by small holes," Phys. Rev. **66**, 163-182 (1944).
2. T.W. Ebessen, H. J. Lezec, H. F. Ghaemi, T. Thio and P. A. Wolff, "Extraordinary optical transmission through subwavelength hole arrays", Nature **391**, 667-669 (1998).
3. R. Gordon, "Light in a subwavelength slit in metal: propagation and reflection", Phys. Rev. B **73**, 153405 (2006).
4. J. Bravo-Abad, L. Martin-Moreno and F.J. Garcia-Vidal, "Transmission properties of a single metallic slit: from the subwavelength regime to the geometrical optical limit", Phys. Rev. E **69**, 026601 (2004).
5. Y. Xie, A. Zakharian, J. Moloney, and M. Mansuripur, "Transmission of light through slit apertures in metallic films", Opt. Express **12**, 6106-6121 (2004).
6. F. Yang and J. R. Sambles, "Resonant transmission of microwaves through a narrow metallic slit", Phys. Rev. Lett. **89**, 063901 (2002).
7. J. R. Suckling., A. P. Hibbins, M. J. Lockyear, T. W. Preist, J. R. Sambles, and C. R Lawrence, "Finite conductance governs the resonance transmission of thin metal slits at microwave frequencies", Phys. Rev. Lett. **92**, 147401 (2004).
8. M. A. Vincenti, M. De Sario, V. Petruzzelli, A. D'Orazio, F. Prudenzano, D. de Ceglia, N. Akozbek, M. J. Bloemer, P. Ashley and M. Scalora, "Enhanced transmission and second harmonic generation from subwavelength slits on metal substrates", Proc. SPIE **6987,** 69870O (2008).
9. Y. Takakura, **"**Optical Resonance in a Narrow Slit in a Thick Metallic Screen", Phys. Rev. Lett. **86**, 5601-5603 (2001).
10. H. Raether, "Surface polaritons on smooth and rough surfaces and on gratings", Springer-Verlag, Berlin (1988).
11. E.D. Palik, "Handbook of Optical Constants of Solids", Academic Press, London-New York (1985).
12. M.A. Vincenti, A. D Orazio, M. G. Cappeddu, Neset Akozbek, M. J. Bloemer, M. Scalora, "Semiconductor-based superlens for sub-wavelength resolution below the diffraction limit at extreme ultraviolet frequencies", Journal of Applied Physics **105**, 103103 (2009).
13. M. Scalora, G. D'Aguanno, N. Mattiucci, M. J. Bloemer, J. W. Haus and A. M. Zheltikov, "Negative refraction of ultrashort electromagnetic pulses", Appl. Phys. B **81**, 393-402 (2005).

## 1. Introduction

Propagation of light through small apertures has fascinated scientist for centuries. As the size of the aperture approaches the wavelength it behaves like a point source. When the aperture size is less than the wavelength the transmission of light is also significantly reduced. Bethe [1] calculated the transmission through sub-wavelength apertures in a perfectly conducting, infinitely thin screen. In this case the transmission is greatly reduced when the size of the aperture is much less than the incident wavelength. The demonstration of extraordinary transmission in metallic sub wavelength array of apertures by Ebessen in 1998 has received renewed interest in this field particularly for its applications in near field optics and nano plasmonics [2]. This peculiar phenomenon has been studied widely from the optical to the microwave regime in an attempt to shed light on the physical mechanisms that contribute it [3-7]. It has been demonstrated both numerically and experimentally that the choice of the geometrical parameters is crucial to the enhanced transmission process [8,9]. On the other hand, the excitation of surface waves inside and outside the holes or apertures is indeed necessary for the process [8]. Several theoretical studies have reported on the transmission and reflection of light through sub-wavelength holes on *perfect electric conductor* (PEC) screens [5, 6, 8], showing that the calculation of the propagation constant and the wavelength of the Fabry-Peròt resonance inside the slit reduces to an arithmetical exercise. However, as predicted by surface plasmon theory [10], the introduction of a negative real part of the relative permittivity of the metal allows the formation of surface modes, leading to efficient coupling of the incoming light inside the sub-wavelength holes imprinted on the substrate. Since a negative value for the permittivity is achieved only if the operating wavelength is tuned below the plasma frequency of the chosen metal, the applications that exploit the extraordinary transmission regime for linear and nonlinear optical applications are mostly limited to the visible and infrared ranges.

As calculations on silver substrates have demonstrated, cavity effects relying on the formation of surface plasmons are indeed simultaneously important in both linear and nonlinear processes like second harmonic generation from apertures on metal substrates [8]. As a representative example, in Fig.1 we plot a two-dimensional map of the transmission from a single slit in a smooth silver substrate as a function of slit width and depth with TM-polarized incident wave tuned at 800nm. This shows that for a given incident wavelength the maximum transmission depends on slit width and substrate thickness. To further elucidate this, in Fig.2 a cross section of the transmittance as a function of depth is shown, when the slit size is fixed as 32 nm. For a real metal, the longitudinal resonance phenomenon makes it possible to achieve the extraordinary transmission regime without the need for additional surface texturing or grating for the excitation of surface waves, as the energy is funneled into the cavity and efficiently re-emitted on the opposite side of the substrate [8].

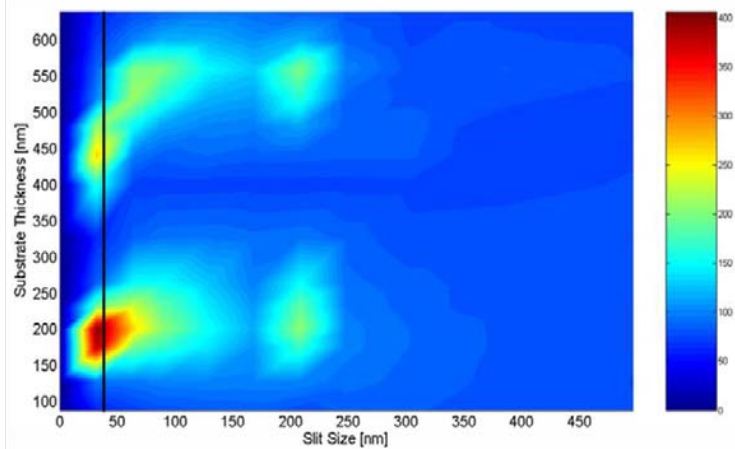

**Fig.1:** Two-dimensional map of the transmission coefficient evaluated by varying the substrate thickness and the aperture size. The presence of two resonances for the same substrate thickness (~200nm) confirms the Fabry-Peròt like behavior of this structure [8].

We now extend the extraordinary transmission to semiconductors. Unlike metals where the plasma frequency is in the UV-visible range, in semiconductors the plasma frequency is usually in the deep UV. For example, the dielectric constant of GaAs and other semiconductors becomes negative just below 270nm [11] (see Fig.3). In this article we theoretically demonstrate that by choosing a proper slit size and GaAs layer thickness the extraordinary transmission regime can be achieved, despite the large absorption in the deep

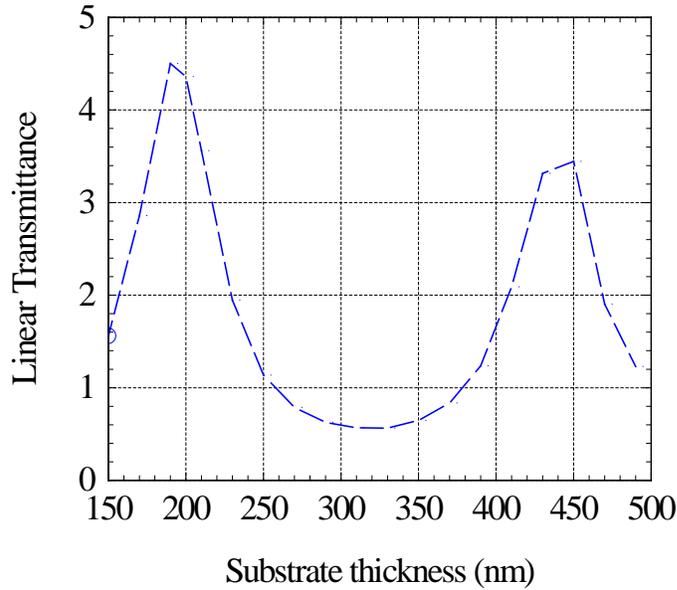

**Fig.2:** Transmittance vs. substrate thickness for a 32nm aperture on a Silver substrate for 800nm incident wavelength. The data corresponds to the transmittance traced by the black line on Fig.1 above.

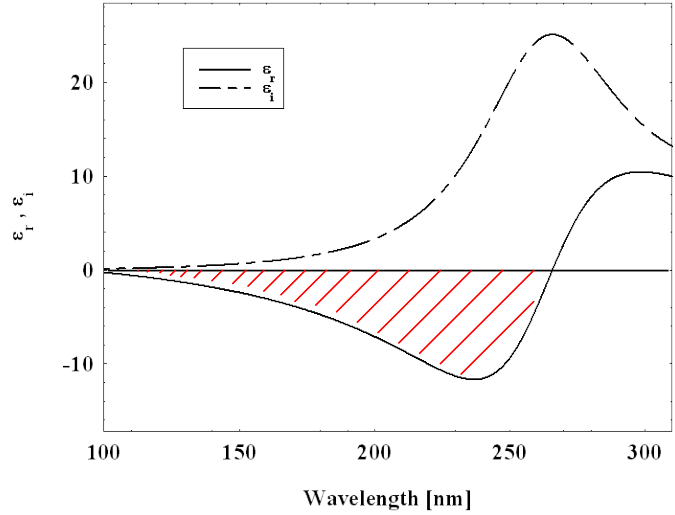

**Fig.3:** Real and imaginary parts of the dielectric permittivity of GaAs below 300nm. The shaded area indicates the range of negative dielectric constant.

UV and soft X-ray ranges [12]. Extraordinary transmission is achieved through a combination of surface waves and field localization inside the nano-slits and may open a new wavelength regime in semiconductors for both linear and nonlinear responses previously not studied. Resonant conditions thus lead to the formation of surface waves and field localization inside nano-slits at wavelengths that are usually deemed inaccessible, that may instead be useful for linear and nonlinear optical applications. In contrast to metals, which do not have an intrinsic quadratic nonlinear term, nonlinear interactions in semiconductors are mostly driven by a $\chi^{(2)}$ contribution, although under the right circumstances surface phenomena may compete and acquire some importance. It follows that the bulk nonlinear response of semiconductors has a different behavior relative to metal apertures [8], as the geometrical parameters of the slit are varied. We will explore the nonlinear optical properties of apertures on semiconductor substrates separately.

## 2. Linear Response of a Single Sub-Wavelength Slit Carved on a GaAs Substrate

We consider a single layer of a GaAs having the linear dispersion profile taken from Palik's handbook of optical constants [11], and having thickness *w* and an aperture of size *a* (Fig. 4). The variation of these two geometrical parameters is examined in order to enhance or maximize linear transmittance. All calculations reported below were performed using two different techniques to make sure the physics is modeled correctly, by solving directly Maxwell's equations using an in-house Finite-Difference Time Domain (FDTD) method and a Time-Domain Fast Fourier Transform Beam Propagation Method (FFT-BPM) [13]. For both numerical methods we considered a Gaussian-shaped incident field tuned at $\lambda = 200$ nm. As depicted in Fig. 5, we considered three different values of *a*, which are smaller than the operating wavelength ($\lambda/12$, $\lambda/8$ and $\lambda/6$), and varied the thickness of the substrate from $w=\lambda/10$ to $w=\lambda$. By normalizing the transmitted field to the energy that impinges on the geometrical area behind the slit, we observed how the resonances form in nearly the same positions by increasing aperture size; at the same time the value of the transmission increases with aperture size, approaching a value of 40% when $a=32$nm and $w=40$nm. It is worth noting that performing the same analysis for wider slits does not alter the longitudinal Fabry-Peròt-like behavior of the nano-cavity, but the process becomes less interesting when the value of *a* becomes comparable to $\lambda$: the influence of aperture size on the enhanced transmission phenomenon is relevant only if the size of the aperture is much smaller than the operating wavelength, allowing the efficient formation of longitudinal surface modes inside the hole. Once we fixed the geometrical parameters to $a=32$nm and $w=40$nm, we varied the incident wavelength to investigate the relationship between the extraordinary transmission regime and the dispersion peculiarities of GaAs. As Fig.6 shows, the single slit system exhibits strong resonant behavior as a function of wavelength. The magnitude of the transmittance (blue curve, left axis) increases as the magnitude of the real part of the permittivity decreases (red curve, right axis), reaching a maximum value of ~102% when $\lambda=240$nm. Fig.6 demonstrates quite clearly that the enhancement of transmission is proportional to $\varepsilon_r$ and somewhat independent of absorption (black curve, right axis). These results prove also that the extraordinary transmission regime is achieved in spite of the large absorption that typically characterizes semiconductors in the UV range: in Fig.6 maximum transmission occurs practically at resonance. As Fig.7 shows, the slit exhibits a Fabry-Peròt extraordinary transmission regime in semiconductors despite large absorption in the UV positions of the maxima and the minima of the transmission spectrum nearly correspond to the theoretically predicted obtained surface plasmon theory. Once the operating wavelength is determined, the geometrical analysis helps to further optimize the device in boosting the transmission well above 100%. For the incident wavelength λ=240nm, the effective index for the surface plasmon that propagates at the GaAs/Air interface is ~1.05, which means that the surface wave propagates at $\lambda_{sp}$~227.83nm. By simply considering the slit as a Fabry-Peròt cavity, one can expect the maxima of the transmission at $\lambda_{sp}/4$ (~57nm) and $3\lambda_{sp}/4$ (~170 nm). At the same time, the minimum of the transmission must occur at $\lambda_{sp}/4$ (~ 114 nm). These values are obtained by neglecting the imaginary part of the dielectric permittivity of the semiconductor, results only in a small shift of the resonant positions by less than 5 nm.

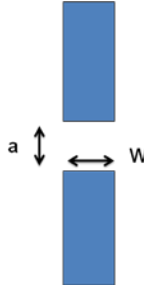

**Fig. 4:** Sketch of the simulated structure: a sub-wavelength slit *a* wide is carved on a GaAs layer *w* thick.

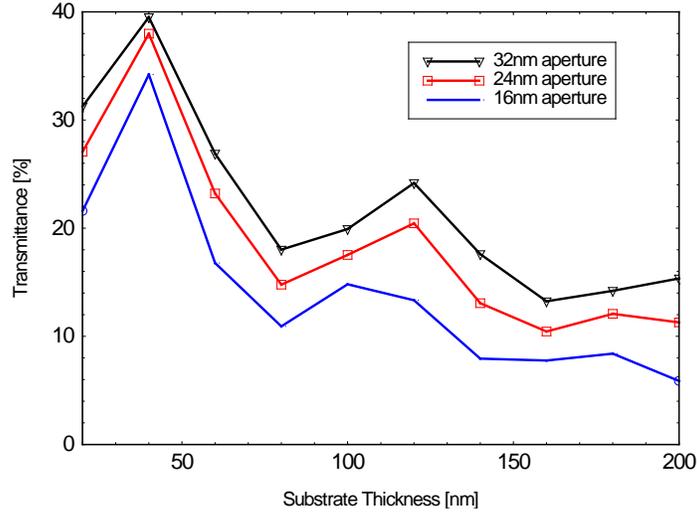

**Fig. 5:** Transmission coefficient of a single slit carved on a GaAs substrate: results are plotted varying the thickness of the substrate layer with slits widths *a* = 16 nm, 24 nm and 32 nm

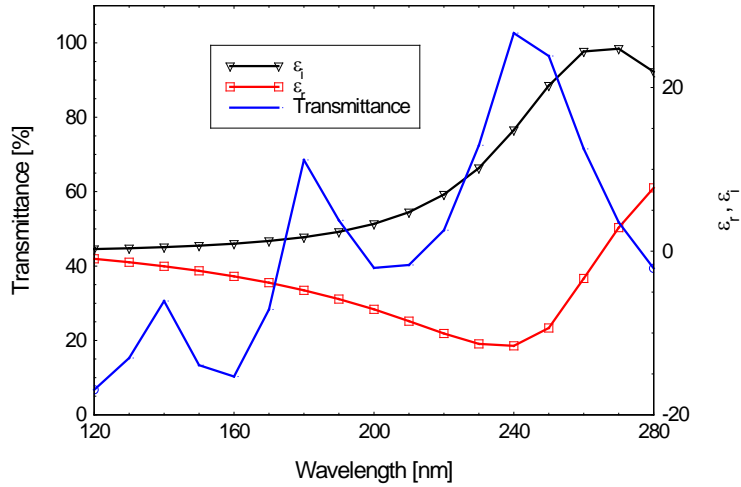

**Fig. 6:** Spectral Response of a single 32 nm aperture carved on a 40 nm thick GaAs substrate (blue curve); the figure reveals a resonant behavior and the dependence of the transmission process on the permittivity values of the semiconductor (red and black curves)

## 3. The contribution of Surface Plasmons to the Enhanced Transmission regime: real substrates versus perfect screens

A nontrivial aspect of the linear response depicted in Fig.6 is the significant difference relative to the same quantity obtained for light propagating through a PEC screen, which is usually reported in the literature as a reference. An accurate analysis of transverse field profiles and localization inside the nano-slit at resonance clarifies the nature of the discrepancies that give rise to the formation of *hot spots* inside a sub-wavelength aperture. The Fabry-Peròt-like behavior of these nano-cavities may be altered significantly by the introduction of a realistic dielectric permittivity: by imposing a PEC condition, which also means that one is neglecting the value of the real part of the permittivity, voids the formation of surface waves and forces the fields to be zero at the surface. These are not realistic conditions for either metals in the visible and near IR ranges, or semiconductors in the UV range. So questions arise about qualitative and quantitative aspects of the interaction compared to the realistic conditions that one should instead examine, and that culminate with significant field enhancements near and around the nano-slit and the enhanced transmission process. Moreover, the smaller the aperture is, the stronger the contribution of surface plasmons will be to the transmitted field and the differences between a real material and a PEC: as depicted in Fig. 8, plotting the transmission value for a slit of variable size in GaAs (blue curve, left axis) and in a PEC (red curve, left axis) is quite evident that the enhancement factor (black curve, right axis) decreases when the aperture size is increased, yielding a value of ~ 1.5 for apertures comparable to the incident wavelength. By looking at Fig. 8 one could also infer that an enhancement of ~ 1.5 is guaranteed by the formation of surface waves on the wall alone, while this enhancement could be boosted significantly when field penetration is allowed and close walls also interact (apertures smaller than 60 nm). As proof of the fact that fields are pushed to assume different shapes inside the nano-cavity thanks to the contribution of the actual permittivity data, in Figs.9 and 10 we report the calculated magnetic fields, the transverse Poynting vector, and electric field for both screen types. These figures highlight how the formation of surface waves inside the slit not only contributes to their shape but also to their intensities, which is more than three times larger for the magnetic field (Figs.9) and two order of magnitude greater in the real material for both transverse components of the Poynting vector and electric field (Figs.10).

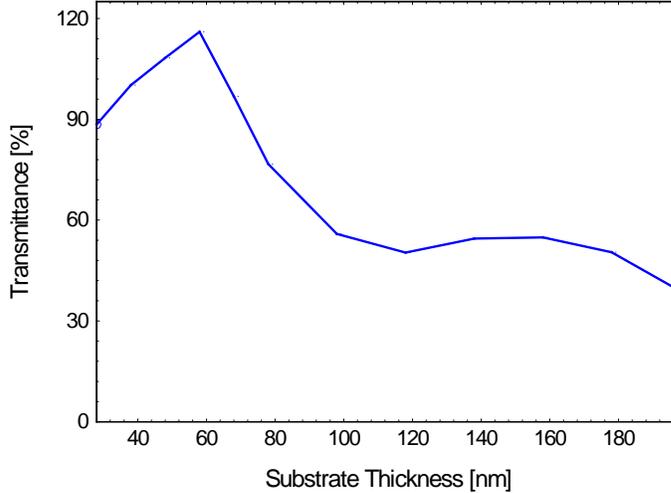

**Fig. 7:** Transmittance vs. substrate thickness for a 32 nm aperture on a GaAs substrate when the incident wavelength is 240 nm.

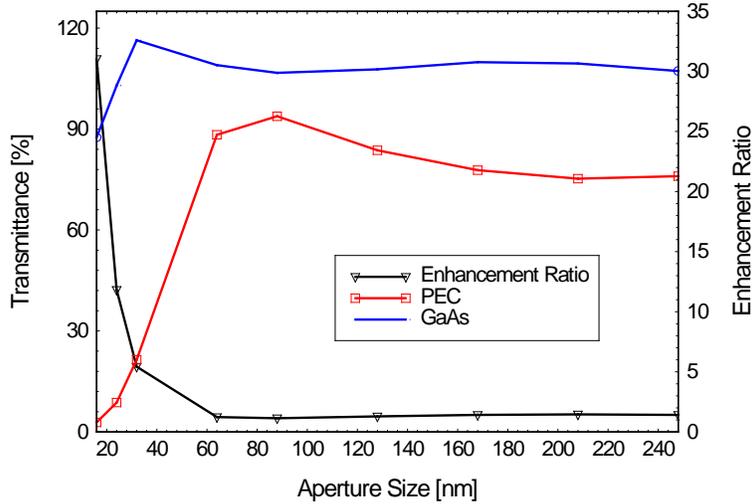

**Fig. 8:** Transmission response for a slit carved on a GaAs (blue cruve, left axis) and a PEC (red curve, left axis) screen; the black curve (right axis) measures the ratio blue/red curves.

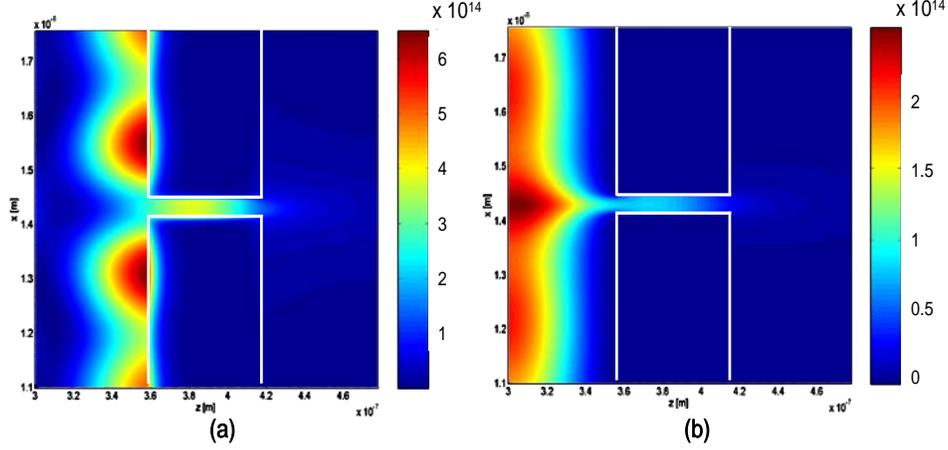

**Fig. 9:** Magnetic field in a 32 nm slit carved on 58 nm of a) GaAs and b) PEC.

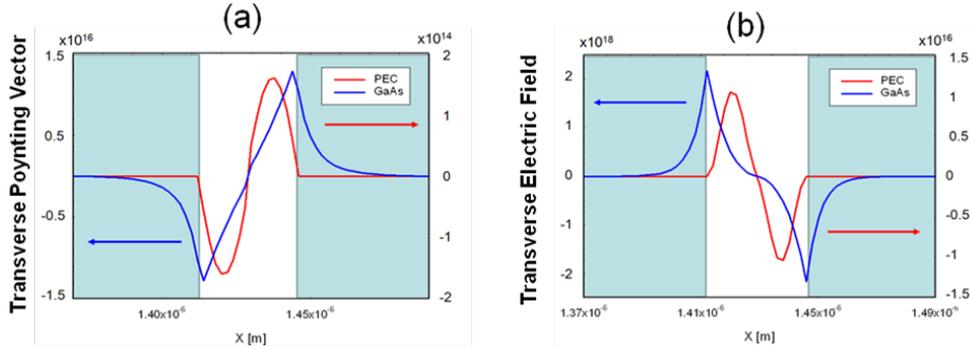

**Fig. 10:** a) Transverse Poynting vector for GaAs and a PEC screen where a 32 nm wide aperture has been carved on a 58 nm substrate. b) Transverse Electric Field for GaAs and PEC screen for the same geometrical sizes as a).

### 4. Multiple Apertures

As it was demonstrated for resonant silver substrates [8], the overall transmittance can be much higher in a multi-slit system thanks to a combination of the formation of surface waves that can constructively interfere between neighboring apertures and properly chosen substrate thickness. However, since slit size and substrate thickness are fixed to maximize the transmission process, for simplicity here we report only the dependence of transmittance as a function of slit separation. As depicted in Fig.11, for λ=240 nm, two similar slits having a center-to-center distance of ~λ/3 (to be contrasted with λ/2 for silver [8]) yields a transmittance of ~147%. This improvement amounts to a *total absolute* transmittance (normalized with respect to the total energy incident on the entire surface) close to 60%, and is comparable with improvements obtained for silver. However, the improvement is not preserved for large number of slits. While silver shows a small increase of the transmission coefficient and saturation beyond a few apertures, the large imaginary part of the dielectric function of GaAs frustrates the transmission process as the number of slits increases (Fig.12).

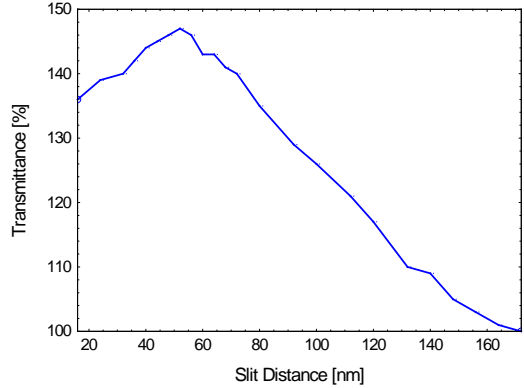

**Fig. 11:** Transmission of a double slit system; a=32nm and the inter-slit pitch varies from 16 to 170nm. Substrate thickness and incident wavelength are 58nm and 240nm, respectively.

### 4. Conclusions

We have demonstrated that it possible to achieve extraordinary transmission of light in the UV and soft X-ray ranges by exploiting the negative value of the dielectric permittivity of semiconductors, in spite of large absorption values. For certain geometrical parameters and excitation wavelengths we demonstrated a transmittance of ~147 % for a double slit system, thanks to the coupling of surface waves inside sub-wavelength sized apertures, also for materials that exhibit large absorption. For GaAs we find a transmission maximum near the absorption resonance, at ~240 nm. Together with the improvement of the linear process we also predict (but will report separately) direct correlations to the enhancement of nonlinear interactions, just as occurs for the metal substrates [8], albeit with different qualitative aspects with respect to the metal, in this case due to the presence of a bulk nonlinearity.

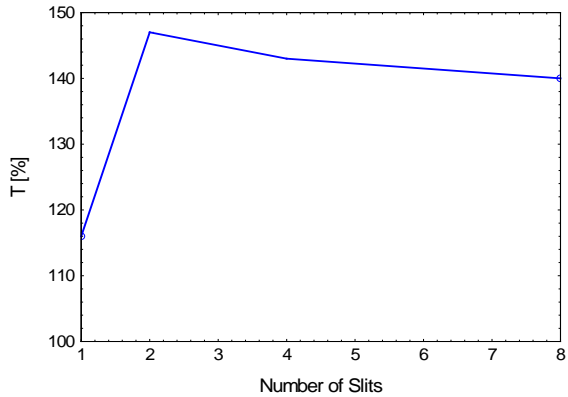

**Fig. 12:** Transmission coefficient as a function of the number of slits. The transmitted power decreases because an absorption resonance characterizes GaAs in this wavelength range.